\documentclass[12pt]{article}

\usepackage{graphicx}
\usepackage{bm}
\usepackage{amsmath}
\usepackage{amssymb}
\usepackage{chicago}

\begin{document}

\title{Dynamics of fixation of advantageous mutations}

\author{Viviane M. de Oliveira\dag  and Paulo R. A. Campos \ddag}

\maketitle

\noindent
Instituto de F\'{\i}sica Gleb Wataghin, Universidade Estadual de Campinas \\
13083-970 Campinas SP, Brazil

\bigskip

\noindent 
Corresponding author:  Paulo R. A. Campos. Phone: 55-19-3788-5373, Fax: 55-19-3788-5376, email: prac@ifi.unicamp.br


\newpage

\begin{abstract}
We investigate the process of fixation of advantageous mutations in an asexual population.
We assume that the effect of each beneficial mutation is exponentially distributed with mean value $\omega_{med}=1/\beta$.
The model also considers that the effect of each new deleterious mutation reduces the fitness of the organism independent on the
previous number of mutations. We use the branching process formulation and also extensive simulations to study the model.
The agreement between the analytical predictions and the simulational data is quite satisfactory. 
Surprisingly, we observe that the dependence of the probability of fixation $P_{fix}$ on the parameter $\omega_{med}$ is precisely
described by a power-law relation, $P_{fix} \sim \omega_{med}^{\gamma}$. The exponent $\gamma$ is an increase function of the 
rate of deleterious mutations $U$, whereas the probability $P_{fix}$ is a decreasing function of $U$.
The mean value $\omega_{fix}$ of the beneficial mutations which reach ultimate fixation depends on $U$ and $\omega_{med}$. 
The ratio $\omega_{fix}/\omega_{med}$ increases as we consider higher values of mutation value $U$ in the region of 
intermediate to large values of $\omega_{med}$, whereas for low $\omega_{med}$ we observe the opposite behavior.
\end{abstract}

\section{Introduction}
The process of adaptation in evolving populations takes place by the continuous production of beneficial mutations and the
ultimate fixation of these variants in the population. This mechanism is relevant not only to improve the adaptation of the organisms
to the environment, but also to prevent that in an environment where a large supply of slightly deleterious mutations persists,
the population presents a continuous decline of the mean population fitness leading to the extinction 
of the whole population. Besides, beneficial mutations have a crucial role to permit the 
adaptation in dynamic environments where drastic changes take place. 
The attempt to understand the dynamics of fixation of beneficial mutations stems from the classical population 
genetics with the pioneer works of Fisher and Haldane \cite{fisher22,fisher30,haldane27}. With the developments in the 
experimental biology and the consequent abundance of data from real biological systems, especially those 
from bacteria \shortcite{visser99,rozen2002,shaver2002} and viruses
populations \shortcite{miralles99,cuevas2002}, it is now possible to better understand the main mechanisms 
underlying the process of adaptation in
these populations and the rhythm at which it occurs. In this sense, recent theoretical developments have
contributed 
to the advances in the field \cite{gerrish98,orr2000,barton95,barton2002,gerrish2001}. For instance, it is known 
that the increase of the supply of beneficial mutations in an asexual population
does not result in a linear response of the evolutionary process 
due to the competition between distinct lineages in order to reach fixation. In an asexual population beneficial mutations
are fixed sequentially and in the case where two or more mutations compete for fixation only one mutation can be kept in the 
population, with the definitive loss of the remaining ones. This process is named clonal interference\shortcite{hill66,campos2003}.

Here, we consider that beneficial mutations are rare events and so they do not compete for fixation with other mutations. 
Nevertheless, deleterious mutations occur at a constant rate $U$. Deleterious mutations affect drastically the dynamics of fixation of
the beneficial mutations. As previously demonstrated \cite{CamposBMB2003,peck94}, the 
probability of fixation $P_{fix}$ of advantageous mutants is
a decrease function of the mutation rate $U$. This is a consequence of the possibility of occurrence of such 
beneficial mutation in a genome with a large amount of segregated deleterious mutations, i.e., in a genome with very low
fitness value.  
In this paper we consider the multiplicative fitness landscape at which the effect of each new deleterious mutation is independent
of other genes.

The paper is organized in the following way: In the next Section we describe the model. 
In Section III we discuss the branching process formulation. In Section IV we show our results, and finally
In Section V we present our conclusions.

\begin{figure*}[t]
\includegraphics[width=12cm,angle=0]{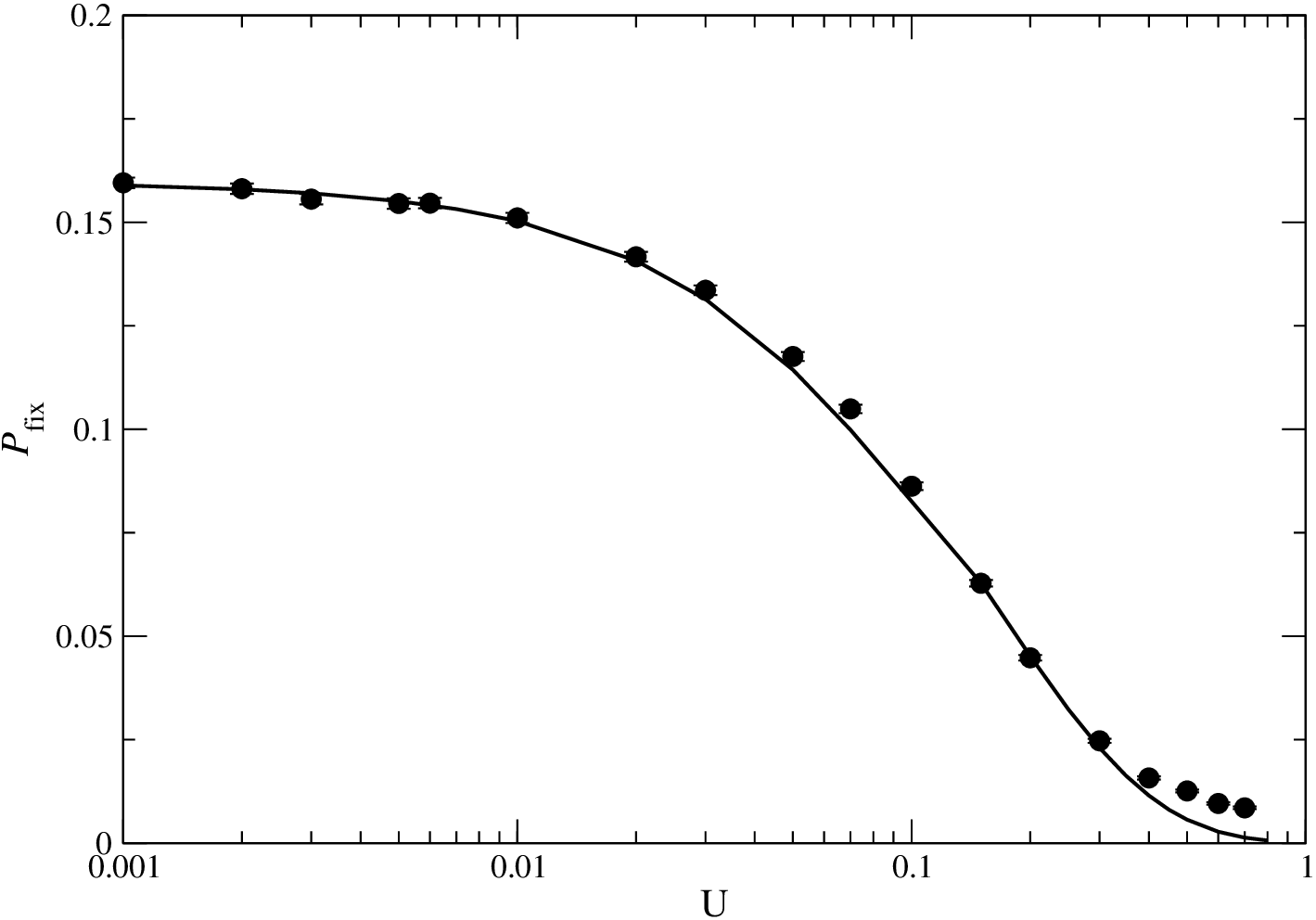}% Here is how to import
\caption{\label{fig:figure1} Probability of ultimate fixation as a function of the mutation rate $U$. 
The parameters are $N=1000$, $\beta=10$ and $s_{d}=0.1$. The data points correspond to the simulation results over $100,000$
runs, and the thick line is the theoretical prediction according to Eq. (\ref{Pfix}).}
\end{figure*}

\section{The evolutionary model}
The population consists of $N$ asexual haploid organisms that evolve according to the Wright-Fisher model. In the model, 
the individuals in generation $t+1$ are direct descendants of the individuals in time $t$. The probability that an organism is descendant
of a particular parent in the previous generation is proportional to the parent's fitness. 
We also assume that the genome of an individual is represented by an
infinitely large sequence of bits $\mathbf{S}=(s_{1},s_{2},\ldots,s_{\infty})$, where the digit $s_{\alpha}$ denotes the state of 
gene $\alpha$, which can take two distinct values $s_{\alpha}=0,1$. The state $s_{\alpha}=0$ means that the digit $\alpha$ 
remains in the original state of the ancestor of the population, whereas the state  $s_{\alpha}=1$
means that the nucleotide $\alpha$ has been hit by a mutation. Because we assume infinitely large genomes, the 
probability of reverse mutations is negligible, i.e., we consider that only transitions of the 
type $s_{\alpha}=0 \rightarrow 1$ can take place.

When a newborn individual arises it acquires the number of mutations present in its parent's genome and an additional
amount of new deleterious mutations $n$ taken from a  Poisson distribution with parameter $U$, 
where $U$ is the mean number of new mutations per individual per generation. The aforesaid model was 
introduced by Kimura and Watterson \cite{kimura64,watterson75} and it is referred to the infinite-sites model.
The fitness of an individual depends on the total number $k$ of mutations in its genome and is given by
\begin{equation}
w_{k}=(1-s_{d})^{k},
\end{equation}
where $s_{d}$ is the cost associated to each deleterious mutation. This case 
corresponds to the multiplicative landscape, where each new mutation reduces the fitness of the organism by the same factor 
\shortcite{CamposSonoda}.

For infinitely large populations, the distribution of frequencies of the class of individuals with $k$ mutations in the 
equilibrium regime, which we denote by
$\bar{C}_{k}$, can be calculated by the following set of equations
\begin{equation}\label{frequencies1}
\bar{C}_{k}=\frac{1}{w_{m}-w_{k}}\sum_{j=m}^{k-1}\frac{U^{k-j}}{(k-j)!}w_{j}\bar{C}_{j} ~~~~~ k>m.
\end{equation}
Using the above expression we can recursively calculate the ratios $\bar{C}_{k}/\bar{C}_{m}$ and estimate $\bar{C}_{m}$ from the
normalization condition $\sum_{K} \bar{C}_{k}=1$ \cite{fontanari}, where $m$ is the index of the class of the fittest individuals
existing in the population. In our analysis we always assume $m=0$.   

In our simulations, the initial population is distributed in different classes of individuals according to the
frequencies of equilibrium, given by Eq. (\ref{frequencies1}). In this case, it is not necessary to wait
the population evolves up to reaching the stationary regime.

In the first generation, we randomly select an individual which acquires an advantageous mutation with
selective effect $s_{b}$ obtained from an exponential distribution   
\begin{equation}\label{exponential}
g(s_{b}) = \beta\exp(-\beta s_{b}),
\end{equation}
which is the expected distribution, as argued by the extreme value theory \cite{gillespie,orr2003}.
Whether this mutation happens in a genotype with $k$ deleterious mutations its adaptation 
value increases by a factor $(1+s_{b})$, i.e.,
\begin{equation}\label{beneficial}
w_{k}=(1+s_{b})(1-s_{d})^{k}.
\end{equation}
The advantageous mutation can be propagated for future generations as soon as the individuals which have acquired it
 replicate. The fitness of those individuals carrying the beneficial mutation takes the same form as in Eq. (\ref{beneficial}.)

In our approach, we consider the beneficial mutation to be fixed when the genotype that has first acquired 
it becomes the most-recent common ancestor of the whole population \shortcite{barton95,barton2002,CamposBMB2003,wilke2003,campos2003}.

\section{Branching Process Formulation}
The theory of branching process \cite{harris63} was first used in the context of population genetics by R. A. 
Fisher \cite{fisher22,fisher30} to study the survival
of the progeny of a mutant gene and random fluctuations in the frequencies of genes. Subsequently, Haldane 
used the theory to investigate the problem of
fixation of an advantageous allele \cite{haldane27}. 

Haldane demonstrated that the probability $\pi$, that a given 
genotype with selective advantage $s$ reaches fixation in a two-allele model, is given by the 
solution of the following equation \cite{haldane27}:
\begin{equation}\label{pi}
1-\pi = e^{-(1+s)\pi} .
\end{equation} 

\begin{figure}[t]
\includegraphics[width=12cm,angle=0]{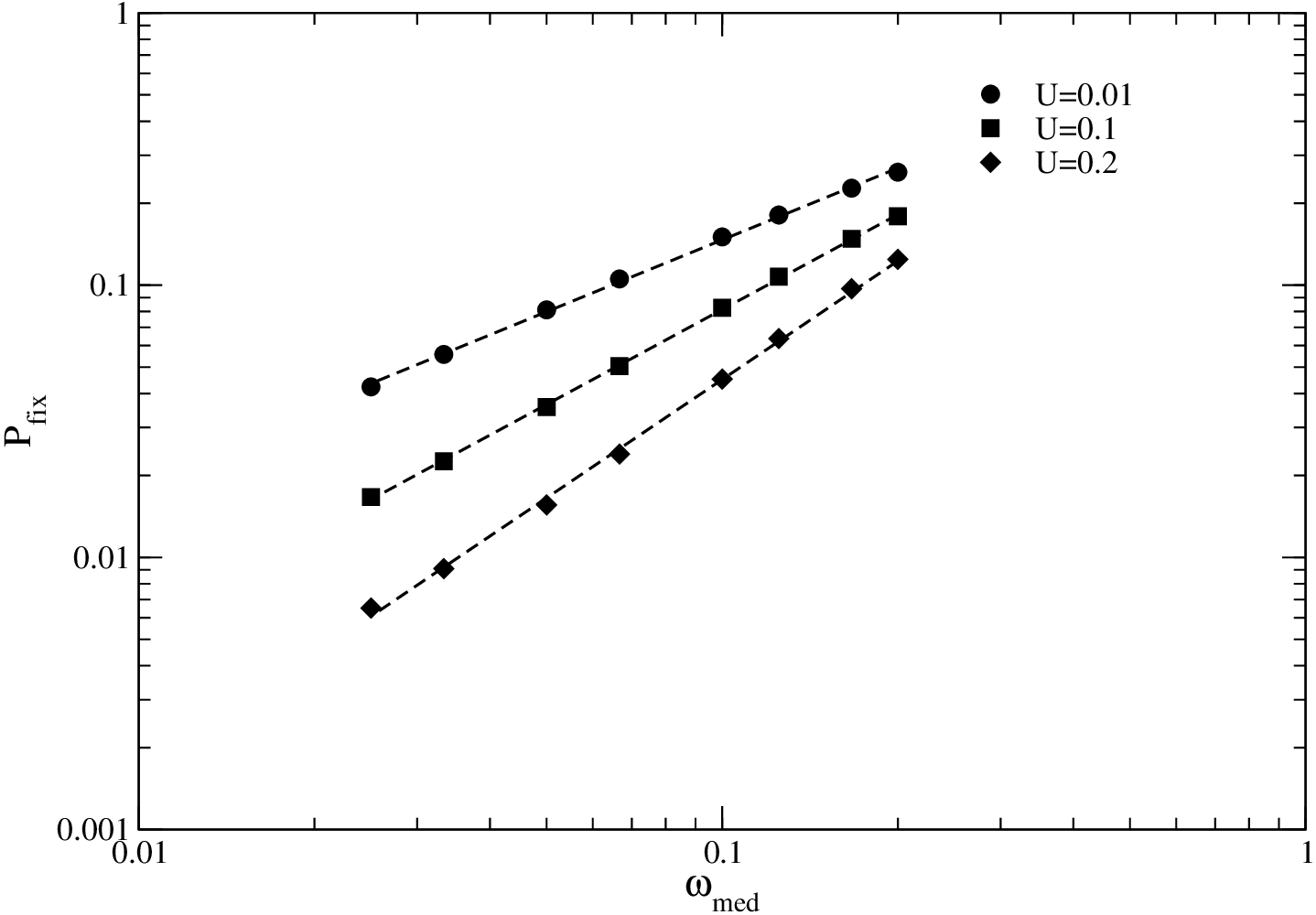}% Here is how to import
\caption{\label{fig:figure2} Probability of fixation $P_{fix}$ as a function of the parameter $\omega_{med}$ for
fixed values of mutation rate $U$. The data points are the theoretical predictions and the dashed-lines are the best fits
which give a power-law distribution. The parameter values are $s_{d}=0.1$ and from top to bottom $U=0.01$, $U=0.1$ and $U=0.2$.}
\end{figure}

\begin{figure}[t]
\includegraphics[width=12cm,angle=0]{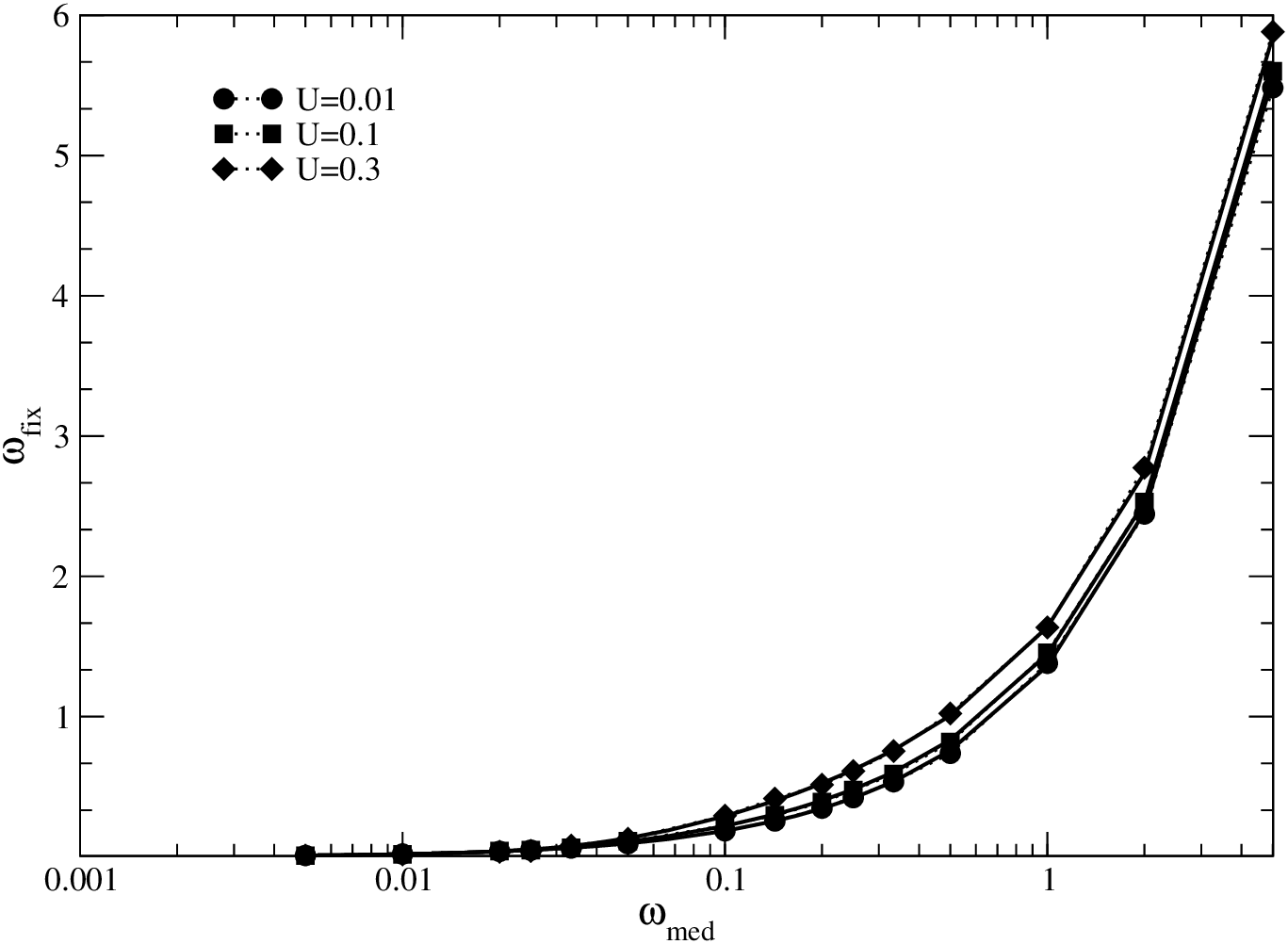}% Here is how to import
\caption{\label{fig:figure3} The mean value of the beneficial effect  of the advantageous mutations that reached fixation $\omega_{fix}$ as
a function of $\omega_{med}$. The parameter values are $N=1000$, $s_{d}=0.1$ and from top to bottom $U=0.3$, $U=0.1$ and $U=0.01$. The
data points correspond to the simulational data whereas the lines are the numerical solutions of Eq. (\ref{wfix}).}
\end{figure}

For small selective values, the solution of this equation yields $\pi (s)\approx 2s$. 

Recently, Barton \cite{barton95,barton2002} extended the use of the branching process formulation to 
heterogeneous genetic background, 
where the individuals can also produce offsprings which are not identical copies of itself.  
In the formulation, the probability $P_{i,t}$ that a beneficial mutation reaches fixation 
when it is present in a single genotype with genetic background $i$ (for instance, $i$ denotes the 
number of mutations) at generation $t$ is obtained by iterating the following set of equations:
\begin{equation}\label{iterative}
(1-P_{i,t-1})=\sum_{j=0}^{\infty}W_{i,j}(1-P_{i,t}^{*})^{j} ~ ,
\end{equation}
where
\begin{equation}
P_{i,t}^{*}=\sum_{k}M_{i,k}P_{k,t}
\end{equation}
is the  probability  that an allele in background $i$ at time $t-1$ would 
get fixed, given that at time {\it t} it is passed to one offspring, and $M_{i,k}$ 
is an element of the mutation matrix that gives the chance that an offspring 
from a parent at background $i$ will be at background $k$. The quantity $W_{i,j}$ is the probability
that an allele in background $i$ contributes with $j$ offsprings to the next generation. If the distribution 
of offsprings is given by a Poisson distribution with mean $\xi_{i}=w_{i}/\bar{w}$ (where $\bar{w}$ is the mean fitness population), 
then
\begin{equation}
W_{i,j} = \frac{\xi_{i}^{j}}{j!}e^{-\xi_{i}},
\end{equation}
and Eq. (\ref{iterative}) is written as
\begin{equation}\label{iterative1}
(1-P_{i,t-1})=\exp \left[-\xi_{i}P_{i,t}^{*} \right].
\end{equation}
The probabilities of fixation correspond to the solution of Eq. (\ref{iterative}) obtained 
in the limit $t \rightarrow \infty$, which we denote by $P_{i}=P_{i,t\rightarrow \infty}$.

\begin{figure*}[t]
\includegraphics[width=12cm,angle=0]{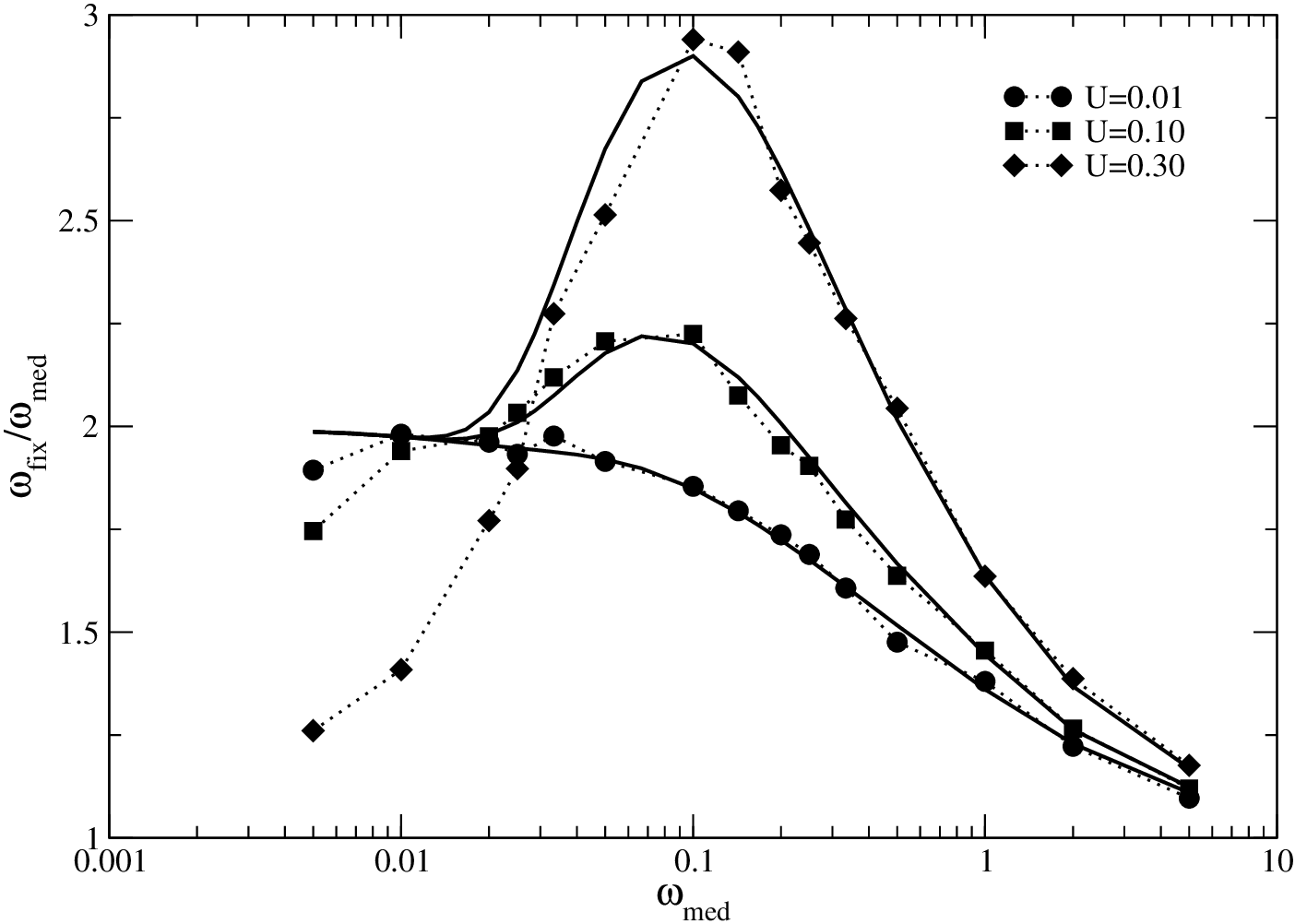}% Here is how to import
\caption{\label{fig:figure4}  The ratio $\omega_{fix}/\omega_{med}$ as a function of $\omega_{med}$. 
The parameter values are $N=1000$, $s_{d}=0.1$ and from top to bottom $U=0.3$, $U=0.1$ and $U=0.01$. The
data points correspond to the simulational data whereas the lines are the numerical solutions of Eq. (\ref{wfix}).}
\end{figure*}

We estimate the probability of fixation of a beneficial mutation with fixed selective effect $s_{b}$, $\Theta_{fix} (s_{b})$, by 
the sum over the distinct genetic backgrounds $i$
\begin{equation}\label{Pfix}
\Theta_{fix} = \sum_{i} P_{i}\bar{C}_{i}.
\end{equation}
The concentrations of the classes $i$, $\bar{C}_{i}$,  also means the chance of occurrence of the beneficial 
mutation in an individual 
in background $i$.

As we consider that the beneficial effect of an advantageous mutation is exponentially distributed 
according to Eq. (\ref{exponential}), the probability of ultimate fixation $P_{fix}$ of a beneficial mutation is then given by
\begin{equation}\label{Pfix}
P_{fix} = \int_{0}^{\infty} g(s_{b})\Theta(s_{b})d s_{b}.
\end{equation}
Another relevant measurement in our statistics is the quantity $\omega_{fix}$, which represents the 
mean value of the beneficial effect of those mutations that have reached ultimate fixation, which we calculate as
\begin{equation}\label{wfix}
\omega_{fix} = \frac{1}{P_{fix}}\int_{0}^{\infty} s_{b}g(s_{b})\Theta(s_{b})d s_{b}.
\end{equation}
Although we only solve Equation (\ref{wfix}) numerically, an analytical approximation can be obtained when we consider small 
values of the parameter $\omega_{med}=1/\beta$ and zero mutation rate. In this case, the probability of fixation $\Theta(s_{b})$ is equal
to the solution for the two-allele model (see Eq. (\ref{pi})), i.e., $\Theta(s_{b}) \approx 2s_{b}$. Substituting the expression $\Theta(s_{b}) = 2s_{b}$ into Eqs. (\ref{Pfix}) and (\ref{wfix}) we get
\begin{equation}
\omega_{fix} = \frac{2}{\beta},
\end{equation}
and the ratio between $\omega_{fix}$ and $\omega_{med}$ yields the value 2.

\section{Results and Discussions}

In this section we present and discuss our results from the simulations and also from the theoretical formulation.

In Figure \ref{fig:figure1} we plot the probability of ultimate fixation as a function of the mutation rate for parameter values
$\beta=10$ and $s_{d}=0.1$. We obtain a good agreement between the simulational data and the theoretical curve obtained 
integrating numerically Eq. (\ref{Pfix}). The simulation results were taken over $100,000$ distinct runs. As expected, we
observe a continous decrease of the probability $P_{fix}$ as we increase the mutation value $U$. For high mutation rate $U$ the 
agreement between the simulations and the theoretical prediction is less satisfactory than those seen 
for small and intermediate values of $U$. This problem occurs due to the occurrence of the
Muller's ratchet phenomenon in finite populations, at which the continuous accumulation of deleterious
mutations leads to loss of the best adapted classes of individuals as the population evolves. Thus, for very high $U$ the
population never reaches the equilibrium regime as supposed in the theoretical formulation \cite{CamposBMB2003}.

In Figure \ref{fig:figure2} we display the probability $P_{fix}$ as a function of the mean value of the distribution of 
selective effects $\omega_{med}$ for fixed values of $U$. The probability $P_{fix}$ is an increase 
function of the parameter $\omega_{med}$
since those mutations with higher selective effect have a greater chance of reaching fixation. We observe that this relation 
is well described by a power-law scaling, i.e., $P_{fix} \sim \omega_{med}^{\gamma}$ where the exponent $\gamma$ is an increase function 
of the mutation rate $U$, whereas $P_{fix}$ decreases with $U$, as we can see in Figure \ref{fig:figure1}.

Figure \ref{fig:figure3} shows the mean value of the selective advantages of those mutations that reached fixation $\omega_{fix}$ plotted
against the mean value $\omega_{med}$. As expected $\omega_{fix}$ increases with the augment of $\omega_{med}$, although we do not
see a linear response. Besides $\omega_{fix}$ increases with the raise of $U$, and in this case $U$ only those mutations 
awarding a large beneficial effect to the individuals have a non-negligible chance to reach fixation. We can better 
understand this scenario in Figure \ref{fig:figure4} where we plot
the ratio $\omega_{fix}/\omega_{med}$ as a function of $\omega_{med}$. From the figure, we see that the ratio $\omega_{fix}/\omega_{med}$
is an increase function of the mutation rate $U$ for about all values of the parameter $\omega_{med}$. Nevertheless, an exception is 
obtained when we consider very small
$\omega_{med}$, where we observe that when $U \rightarrow 0$ the ratio $\omega_{fix}/\omega_{med} \rightarrow 2$. Moreover, we 
witness that for intermediate to high values of $U$ the ratio  $\omega_{fix}/\omega_{med}$ is optimized around $\omega_{c} \approx 0.1$.

\section{Conclusions}

We have studied the dynamics of fixation of advantageous mutants in the multiplicative 
fitness landscape.
We have investigated the problem by means of extensive simulations and also by a theoretical approach where we use the
branching process formulation introduced by Haldane and extended by Barton to the case of 
heterogeneous genetic background.
The simulation results are in very good accordance with the theoretical predictions. 
The beneficial effect of these advantageous mutations is assumed to be exponentially distributed with mean value $\omega_{med}$.
At once, we witness that the continuous supply of deleterious mutations to the population reduces the chance of fixation
of the beneficial variants in the population. We have noticed that the relation between the probability of ultimate fixation
$P_{fix}$ and the parameter $\omega_{med}$ obeys a power-law scaling like $P_{fix} \sim \omega_{med}^{\gamma}$, where the exponent
$\gamma$ depends on the mutation rate $U$. Surprisingly, we have observed that the ratio $\omega_{fix}/\omega_{med}$ between 
the selective advantages of those mutations
that reached fixation and the mean value of the distribution is optimized around the 
critical value $\omega_{med}=\omega_{c} = 0.1$ for intermediate
to high values of $U$. In this range of $U$, we have also witnessed that an increase of $U$ also means a higher 
value of $\omega_{fix}/\omega_{med}$. For very low values of $\omega_{med}$, the branching process theory 
fails to predict $\omega_{fix}/\omega_{med}$ when we consider large values of $U$: while the simulations 
show that the ratio $\omega_{fix}/\omega_{med}$ goes to $1$, which corresponds to a random stochastic 
regime, the theoretical analysis gives $\omega_{fix}/\omega_{med}=2$, which is the same 
value attained when we have mutation rate $U=0$. Actually, a smaller value of $U$ means that 
the random stochastic regime will be attained at smaller values of $\omega_{med}$. 

\section*{Acknowledgments}
VMO and PRAC are supported by Funda\c{c}\~ao de Amparo \`a 
Pesquisa do Estado de S\~ao Paulo under Proj. No. 03/00182-0.

\end{document}